\documentclass{article}
\usepackage{spconf}

\usepackage[ left=1in, top=1in, right=1in, bottom=1in]{geometry}
\usepackage{graphicx,bm,colonequals,amsmath,amssymb,url,xcolor,bbm}
\usepackage{array,tabularx,multirow}
\usepackage{enumitem,MnSymbol}
\usepackage[font={footnotesize}]{subcaption}
\usepackage[utf8]{inputenc}
\usepackage{soul} 
\usepackage{placeins} 
\usepackage[normalem]{ulem}
\usepackage{float}
\usepackage{algorithm, algorithmic}
\usepackage{amssymb}
\usepackage{pifont}

\graphicspath{{plots/}}

\usepackage{hyperref}
\usepackage[authoryear]{natbib}

\usepackage{mathtools}
\mathtoolsset{showonlyrefs} 

\bibpunct[, ]{(}{)}{;}{a}{,}{,}
   
\usepackage{amsthm}
\usepackage{array,framed}

\makeatletter
\renewcommand{\paragraph}{%
  \@startsection{paragraph}{4}%
  {\z@}{2ex \@plus 1ex \@minus .2ex}{-1em}%
  {\normalfont\normalsize\bfseries}%
}
\makeatother

\usepackage{etoolbox}
\makeatletter
\preto{\@verbatim}{\topsep=0pt \partopsep=0pt }
\makeatother

\DeclareMathAlphabet\mathbfcal{OMS}{cmsy}{b}{n}

\newcommand{\cmark}{\ding{51}} 
\newcommand{\xmark}{\ding{55}} 

\usepackage{comment}
\begin{document}
\title{Flexible survival regression with variable selection for heterogeneous population}
%
\name{Abhishek Mandal$^{\star\dagger}$,  Abhisek Chakraborty$^{\star\dagger\dagger}$\thanks{Both the authors contributed equally. There was no external or internal funding for this work.}}
\address{$^{\dagger}$Department of Statistics, Florida State University, Tallahassee, FL 32306-4330, USA\\
$^{\dagger\dagger}$Department of Statistics, Texas A$\&$M University, 3143, College Station, TX 77843, USA}
%
%
%
%
\maketitle
\begin{abstract}
Survival regression is widely used to model time-to-events data, to explore how covariates may influence the occurrence of events. Modern datasets often encompass a vast number of covariates across many subjects, with only a subset of the covariates significantly affecting survival. Additionally, subjects often belong to an unknown number of latent groups, where covariate effects on survival differ significantly across groups. The proposed methodology addresses both challenges by simultaneously identifying the latent sub-groups in the heterogeneous population and evaluating covariate significance within each sub-group. This approach is shown to enhance the predictive accuracy for time-to-event outcomes, via uncovering varying risk profiles within the underlying heterogeneous population and is thereby helpful to device targeted disease management strategies.
\end{abstract}
\begin{keywords}
non-parametric Bayes; non-local prior; non-homogeneous population; proportional hazard. 
\end{keywords}

\section{Introduction}\label{sec:intro}
Survival times for patients record the time to death, disease progression, study termination, or loss to follow-up. In the case of study termination or loss to follow-up, the survival time is said to be censored. The functional dependence of the survival times on covariates of interest is often modeled via the hazard function
$h(t \mid \mathbf{x}) = \lim_{\Delta t \to 0} \frac{1}{\Delta t} P(t \leq T \leq t + \Delta t \mid T \geq t, \mathbf{x})$, that represents the instantaneous rate of death at time \( t \), given patient covariates \(\mathbf{x}\) \citep{cox1984analysis,  rossi2014bayesian}. One pertinent challenge in modeling such time-to-events data is identification of features that have a significant influence on survival outcomes. To that end, frequentist penalized likelihood methods such as LASSO \citep{LASSO}, adaptive LASSO \citep{zhang2007adaptive}, Elastic-net \citep{simon2011regularization}, Dantzig selector \citep{antoniadis2010dantzig}, and SCAD \citep{fan2002variable} have gained widespread adoption.
Additionally, Bayesian methodologies leveraging Gaussian priors \citep{ibrahim1999bayesian}, g-priors \citep{held2016objective},  spike and slab priors \citep{tang2017spikeandslab}, mixture of non-local priors \citep{nikooienejad2020bayesian}, horse-shoe priors \citep{maity2020bayesian} have been explored extensively too. Computationally efficient variational Bayes approaches \citep{komodromos2020variational}  have also been proposed. 

Another key practical issue arises when the population of interest consists of an unknown number of latent groups \cite{foster2011subgroup, lipkovich2017tutorial}, and the covariate effects on survival times vary differently in subjects belonging to different subgroups. Detection of such latent sub-groups and sub-group-specific variable significance are instrumental in understanding varying risk profiles in subjects belonging to heterogeneous populations. 

In this article, we propose a Dirichlet process mixture \citep{ferguson1973bayesian} of  Cox's proportional hazard models \citep{cox1984analysis}, equipped with a mixture of spike and non-local slab priors \citep{Johnson2010} for variable selection. This enables us to adaptively determine the number of latent sub-groups from data and identify the sub-groups. We also infer the sub-group specific covariate significance on survival times. The proposed methodology is supplemented with an efficient posterior computation scheme via exploiting the stick-breaking representation of the Dirichlet process \citep{sethuraman1994constructive}, adapting the sure independence screening technique for variable selection \citep{FanLV}, and finally considering convenient Laplace approximations of required marginal likelihoods \citep{tierney1986accurate}. The advantages of the proposed methodology over Bayesian Cox's proportional hazard model \citep{nikooienejad2020bayesian}, in terms of estimation accuracy of regression parameters and variable selection accuracy, are demonstrated via numerical studies. Additionally, the key benefits of the sub-group identification and sub-group specific variable selection are showcased in terms of improved predictive accuracy in publicly available time-to-event data sets on lung cancer and diabetic retinopathy \citep{survival}.

\section{Proposed methodology}
\noindent\textbf{Preliminaries.}
Let ${\bf X}$ be the $n\times p$ design matrix obtained by stacking the $n$ patient covariate vectors. 
The proportional hazard model take the form  $h(t\mid\mathbf{x}_i) = h_0(t)\Phi(\mathbf{x}_i),$
where $h_0(t)$ denotes the baseline hazard function. We additionally need to assume
the identifiability constraint of $\Phi(\mathbf{0})=1$.   The Cox proportional hazards model \citep{cox1984analysis} is then defined by setting $\Phi(\mathbf{x}_i) = \exp\{\mathbf{x}_i^T\mathbf{\beta}\}$, so that $h(t\mid\mathbf{x}_i) = h_0(t)\exp\{\mathbf{x}_i^T\mathbf{\beta}\},$ where $\mathbf{\beta}$ is a $p\times 1$ vector of regression coefficients. An important feature of the proportional hazards model is that it produces a partial likelihood function that is independent of the baseline hazard function. Nonetheless, the baseline hazard function is essential for predicting survival times and can be estimated non-parametrically \cite{nelson1969hazard, aalen1978nonparametric}.  Further, for each subject $\{i\in\{1,2,\ldots, n\}$, suppose \( t_i \) be the observed time, and \( \delta_i \) be the censoring indicator, such that \( \delta_i = 1 \) if the event is observed and \( \delta_i = 0 \) if the time is censored. Then, the complete likelihood function for the Cox model takes the form $\mbox{L}(\boldsymbol{\beta}\mid \mathbf{x}) = \prod_{i=1}^{n} \left[ \left( h_0(t_i) \exp(\mathbf{x}_i^\top \boldsymbol{\beta}) \right)^{\delta_i} \mbox{S}(t_i | \mathbf{x}_i) \right]$, where $\mbox{S}(t | \mathbf{x}) =\exp\left( - \mbox{H}_0(t) \exp(\mathbf{x}^\top \boldsymbol{\beta}) \right)$ is the survival function,
and \( \mbox{H}_0(t) = \int_0^t h_0(u) \, du \) is the cumulative baseline hazard function. For the simplicity of exposition, we assume $h_0(t) = 1$ throughout the article, but the proposed methodology easily generalises for an unknown baseline hazard function.

\noindent\textbf{Motivation.} We recall that the proposed approach is motivated by two practical considerations. Firstly, numerous modern datasets frequently encompass measurements on several variables across hundreds of subjects. From a scientific standpoint, it is plausible that only a relatively small subset of these variables significantly influence survival. This suggests that the majority of the elements in the vector $\mathbf{\beta}$ are either small or nearly zero. The primary challenge, therefore, is to identify the covariates with nonzero coefficients, or equivalently, those covariates that have the most substantial impact on determining the survival outcome. Secondly, it is often the case that there are an unknown number of latent groups among the subjects, and the influence of covariates on survival varies across these different latent groups. Therefore, it is essential to identify the inherent clusters among the subjects and determine the importance of variables in relation to survival outcomes in a cluster-specific manner. 
\vspace{-8mm}
\subsection{Model and prior specification}
With these motivations in mind, we present the first part of our hierarchical specification, utilizing the stick-breaking representation of the Dirichlet process mixture \citep{ferguson1973bayesian, sethuraman1994constructive} of Cox proportional hazard model as follows
\begin{align}
& v_j \stackrel{\text{i.i.d}}{\sim}\mbox{Beta}(1, \alpha),\ \pi_j = v_j\prod_{l=1}^{j-1}(1-v_l), \sum_{j=1}^\infty \pi_j = 1,\notag\\
\end{align}
\begin{align}\label{eqn:model1}
& \beta_k\mid \mbox{G} \sim \mbox{G} \equiv\sum_{j=1}^{\infty} \pi_k \delta_{\beta_j}(\cdot),\notag\\
& z_i \stackrel{\text{ind}}{\sim} \mbox{Categorical}(1, 2,\ldots, \mid \pi_1, \pi_2,\ldots),\notag\\
& h(t_i\mid\mathbf{x}_i, z_i = k) =\exp\{\mathbf{x}_i^T\mathbf{\beta}_k\},\ i = 1,\ldots, n,
\end{align}
where $\alpha>0$ is the concentration parameter. Like earlier, \( t_i \) is the observed time, and \( \delta_i \) be the censoring indicator, such that \( \delta_i = 1 \) if the event is observed, and \( \delta_i = 0 \) if the time is censored. The remainder of the specification consists of specifying priors on the component-specific model size and the regression parameters $\{\mathbf{\beta}_k\}_{k=1}^\infty$.  

To that end, we next discuss the prior on the component-specific model space. The event that the $j$-th variable enters the survival model $\mathbf{m}_k$ in $k$-th group, i.e, $\{j\in \mathbf{m}_k\},\ j= 1,\ldots, p$, is modeled via independent Bernoulli trials. In particular, if we assume $\mbox{P}(j\in \mathbf{m}_k) = \theta\in(0, 1)$, then the resulting marginal probability for the model $\mathbf{m}_k$ becomes $\int_{0}^1 \theta^k (1-\theta)^{p-1-|\mathbf{m}_k|}\ \pi(\theta) d\theta$, where $|\cdot|$ measures the cardinally of a set, and $\pi(\theta)$ is the prior on $\theta$. A completely conjugate choice of $\pi(\theta)$ would be $\beta(a, b)$, which yields 
\begin{align}\label{eqn:model2}
   p(\mathbf{m}_k) 
   = \frac{\beta(a + |\mathbf{m}_k| ,\ b + p-1-|\mathbf{m}_k|)}{\beta(a, b)}, 
\end{align}
for $k\in\{1, 2,\ldots\}$. We set $a=1$ and $b = 1$ to ensure that we do not necessarily assign a low prior probability to a model with many variables.  Next, we focus on the prior specification on the regression coefficients $\beta_j,\ j = 1, \ldots, p$. We put a two-component mixture that has a spike at zero and a non-local slab density. 
In particular, we consider  non-local inverse moment densities \citep{Johnson2010, Johnson2012, Shin2018} 
\begin{align}\label{eqn:model3}
&\pi(u) \propto u^{-(r+1)}\exp{\bigg(-\frac{\tau}{u^2}\bigg)},\quad  u\in\mathbf{R}, 
\end{align}
where $(\tau, r)$ are  prior hyper-parameters. While the hyperparameter $r$ is comparable to the shape parameter in the inverse gamma distribution and controls the tail behaviour of the density, the hyperparameter $\tau$ represents a scale parameter that affects the dispersion of the prior around zero. Notably, the possibility of a large number of variables in the model is not penalised by this prior, in contrast to the majority of penalized likelihood approaches. Consequently, it does not always impose harsh penalties on large models. We also prescribe the default values $r = 1$ and $\tau = 0.25$, which work favourably across the numerical studies. Thus, \eqref{eqn:model1},   \eqref{eqn:model2}, and \eqref{eqn:model3} complete the hierarchical specification of the data generative model. The  corresponding complete likelihood function takes the form  $\mbox{L}(\{\boldsymbol{\beta}_k\}_{k=1}^K\mid \mathbf{x})=$
\begin{align}
\int\prod_{i=1}^{n} \left[\prod_{k=1}^{\infty} \left\{ \left(  \exp(\mathbf{x}_i^\top \boldsymbol{\beta}_k) \right)^{\delta_i} \mbox{S}(t_i | \mathbf{x}_i) \right\}^{1(z_i=k)}\pi(\mathbf{\beta}_k)\right] d\mbox{G},
\end{align}
where $\mbox{S}(t | \mathbf{x}) =\exp\left( - t\times \exp(\mathbf{x}^\top \boldsymbol{\beta}) \right)$ is the survival function. To facilitate the clarity of exposition, we assume a constant baseline hazard function, \( h_0(t) = 1 \), throughout the article. However, the proposed methodology readily extends to the case of an unknown baseline hazard function, which can be accommodated by first performing a preliminary non-parametric estimation of the baseline hazard. 

Note that we not only identify the latent groups within the population and infer the group-specific parameters but also do so without assuming prior knowledge of the number of clusters. One may estimate the number of clusters a-priori using various selection criteria and subsequently estimate the clustering configuration. However, ignoring the uncertainty in the first stage may lead to erroneous clustering. The proposed model-based approach enables us to estimate the number of clusters and clustering configuration simultaneously.

\subsection{Posterior inference}
Let $[\theta\mid\cdot]$ denote the full conditional distribution of any parameter $\theta$ given others. Let $K_{\rm max}$ denote the maximum number of clusters permitted. Sampling from the joint posterior of the parameters interest involves cyclically sampling from the full conditional distribution of each parameter given others as follows.

\noindent\textbf{Step 1.} The full conditional distribution of $[\pi\mid\cdot]$ is $\mbox{Dir}(\alpha/K_{\rm max} +n_1,\ldots, \alpha/K_{\rm max} +n_k)$, where $n_k = \sum_{i=1}^n 1(z_i = k),\ k=1,2, \ldots, K_{\rm max}$. The hyper-parameter $\alpha$ is set at $0.1$. Alternatively, one may put a prior on $\alpha$ and iteratively update it. The specific choice works well in the examples we considered.

\noindent{\bf Step 2.} 
The full conditional distribution of clustering  indices $\mathbf{z} = (z_1,\ldots, z_n)^{\prime}$ is
$\mbox{P}[z_i = k\mid \cdot]\ \propto\ \left\{ \left(  \exp(\mathbf{x}_i^\top \boldsymbol{\beta}_k) \right)^{\delta_i} \mbox{S}(t_i | \mathbf{x}_i) \right\}^{1(z_i=k)}\pi(\mathbf{\beta}_k),$
where $k$ takes values in $\{1,2, \ldots, K_{\rm max}\}$.

\noindent{\bf Step 3.} For $k=1, 2,\ldots, K_{\rm max}$, sampling from the full conditional distribution of the component-specific regression coefficients $\beta_k=(\beta_{k1},\ldots, \beta_{kp})^\prime$ presents the most significant computational bottleneck. Recall that a model $\mathbf{m}_k$ simply denotes all variables with non-zero regression coefficients.  Consequently, we equivalently consider exploring the space of all possible models. In principle, one can  compute   the maximum a posteriori model by solving 
$\hat{\mathbf{m}_k} = \mbox{argmax}_{\mathbf{m}_k\in\Gamma} [\mathbf{m}_k\mid\cdot]$, where $\Gamma$ is the space of all possible models. But, to alleviate the computational challenges in exploring all models in $\Gamma$,   we adapt ideas from Iterative Sure Independence Screening \citep{FanLV}  to compute the mode of $[\mathbf{m}_k\mid\cdot]$. In particular, at each stage of an iteration, we only consider those variables which have a large correlation with the residuals of the current model. To make it precise, at iteration $j+1$, suppose $\mathbf{m}_{k, j}$ is the current model, $r_{\mathbf{m}_{k, j}}$ is the residual from the model $\mathbf{m}_{k,j}$, and $\mathbf{X}_{k,j}$ is the shard of the design matrix corresponding to the observations in the $k$-th cluster. Then, we restrict attention to variables for which $|r_{\mathbf{m}_{k,j}}^{\prime} \mathbf{X}_{k,j}|$ is large. To that end, suppose $S_{\mathbf{m}_{k,j}}$ be the union of variables in $\mathbf{m}_{k,j}$ and the top $M$ variables obtained by screening using the residuals from the model $\mathbf{m}_{k,j}$. Then, we define  $\Gamma_{\rm screen}^{+}(\mathbf{m}_{k,j}) = \{\mathbf{m}_{k,j}\cup\{v\}: v\in\mathbf{m}_{k,j}^{c}\cap S_{\mathbf{m}_{k,j}}\}$ and $\Gamma^{-}(\mathbf{m}_{k,j}) = \{\mathbf{m}_{k,j}\setminus\{v\}: v\in\mathbf{m}_{k,j}\}$. Finally, the screened neighborhood of model $\mathbf{m}_{k,j}$ is defined as $\mbox{N}_{\rm screen}(\mathbf{m}_{k,j}) =\{ \Gamma_{\rm screen}^{+}(\mathbf{m}_{k,j}),\ \Gamma^{-}(\mathbf{m}_{k,j})\}$. This dramatically reduces the computational burden of computing the posterior mode of $[\mathbf{m}_k\mid\cdot]$, since we only need to explore $\mbox{N}_{\rm screen}(\mathbf{m}_{k,j})$, instead of the $\Gamma$ . Refer to Algorithm \ref{alg:s5} for details. 

{\footnotesize
\begin{algorithm}
\caption{Simplified shotgun stochastic search algorithm with screening \citep{FanLV, Shin2018, Amir2016}  in the $k$-th cluster, $k=1,\ldots, K_{\max}$.}\label{alg:s5}
$\mathbf{(1)}$ Choose an initial model $\mathbf{m}_{k, j}^{(1)}$, a set $\mathbf{S}_{\mathbf{m}_{k,j}^{(1)}}$ based on $\mathbf{m}_{k, j}^{(1)}$, and the total number of iterations $N$.\\
$\mathbf{(2)}$ Calculate $\pi(\mathbf{m}_k\mid\mathbf{t}_k)$ for all $\mathbf{m}_k\in \mbox{N}_{\rm screen}(\mathbf{m}_{k,j}^{(1)})$.\\
$\mathbf{(3)}$ 
For $i\in\{1, \ldots, N-1\}$,\\
(a) select $\mathbf{m}_k^{-}$ and $\mathbf{m}_{k}^{+}$ from $\Gamma^{-}(\mathbf{m}_{k, j}^{(i)}), \ \Gamma^{+}_{\rm screen}(\mathbf{m}_{k, j}^{(i)})$ with probabilities $\pi(\cdot\mid\mathbf{t}_k)$ calculated in step (a).\\
(b) Then, select $\mathbf{m}_{k, j}^{(i+1)}$ from $\{\mathbf{m}_{k}^{-}$,\ $\ \mathbf{m}_{k}^{+}\}$ with probabilities $\{\pi(\mathbf{m}_{k}^{-}\mid\mathbf{t}_k)$, 
$\pi(\mathbf{m}_{k}^{+}\mid\mathbf{t}_k)$\}.
\\
(c) Then, update the set of variables considered to $\mathbf{S}_{\mathbf{m}^{(i+1)}_{k, j}}$ to be union of variables in $\mathbf{m}^{(i+1)}_{k, j}$ and top $M$ variables according to $\{|r_{\mathbf{m}^{(i+1)}_{k, j}}^{\prime} \mathbf{X}_{k,j}|, j = 1,\ldots, p\}$.\\
\end{algorithm}
}

\noindent\textbf{Marginal likelihood in Algorithm \ref{alg:s5}.}
The posterior probabilities of the models are not available in closed form. We estimate the posterior  probabilities using Laplace approximations given by $\pi(\mathbf{m}_k\mid \mathbf{t}_k) = (2\pi)^{|\hat{\mathbf{m}}_k|/2}\ \pi(\mathbf{t}_k\mid \hat{\mathbf{m}_k})\ \pi(\hat{\mathbf{m}_k})\ |G(\hat{\mathbf{m}_k})|^{-1/2}$, where $\mathbf{t}_k$ is the shard of survival times of the individuals in the $k$-th cluster, $\hat{\mathbf{m}_k}$ is MAP of ${\mathbf{m}_k}$, $ G(\mathbf{m}_k)$ is the Hessian of the negative log posterior density
$g( \mathbf{m}_k) = -\log\pi(\mathbf{t}_k\mid \mathbf{m}_k) - \log\pi( \mathbf{m}_k)$,
evaluated at $\hat{\mathbf{m}_k}$. 
We use the limited memory version of L-BFGS  to find the maximum a posteriori model. This completes our posterior computation scheme. 

\section{Performance Evaluation }\label{sec:performance}
\noindent\textbf{Experiments.}
We aim to compare the performance of the proposed method with the Bayesian Cox proportional hazard model with no latent groups.  The covariates $x_1,\ldots,x_p$ were generated from multivariate Gaussian distributions with component-wise mean $0$ and variance $1$, and $\mbox{Cor}(x_i,x_j)= \rho$ for $i\neq j \in \{1,\ldots,p\}$. We set $p=40$ and vary $\rho\in\{0.25, 0.5\}$. There are two latent groups, i.e. $K_{\rm true} = 2$, each of size $n_1 = n_2 \in\{100, 200\}$. The size of the true model in each group is $6$, with the nonzero regression coefficients  $\beta_1,\ldots,\beta_6$ sampled from $\mbox{Uniform}(0, 1)$ in group 1 and from $\mbox{Uniform}(25, 26)$ in group 2. The censoring rate $p$ is varied in $\{5\%, 10\%\}$. Each simulation case was then evaluated via repeated simulations.

\begin{table}[!htb]
\caption{\textbf{Variable selection performance comparison.} For varying $\rho\in\{0.25, 0.5\}$ and censoring rate $p\in\{5\%, 10\%\}$.}\label{tab:performance1}
\small\setlength\tabcolsep{11pt}
\scalebox{0.5}{
  \begin{tabular}{ccccccccc}
    \hline
      \multicolumn{1}{c}{} &
      \multicolumn{1}{c}{} &
      \multicolumn{1}{c}{} &
      \multicolumn{3}{c}{$n_1 = n_2= 100$} &
      \multicolumn{3}{c}{$n_1 = n_2= 200$}\\
    \hline
     \multicolumn{1}{c}{$\rho$ } &
     \multicolumn{1}{c}{$p$ } &
      \multicolumn{1}{c}{Method } &
      \multicolumn{1}{c}{Sensitivity} &
      \multicolumn{1}{c}{Specificity} &
      \multicolumn{1}{c}{FDR} &
      \multicolumn{1}{c}{Sensitivity} &
      \multicolumn{1}{c}{Specificity}&
      \multicolumn{1}{c}{FDR} \\
    \hline
0.25  & 5&Ours &1.00 &1.00  & 0.00& 1.00&1.00 &  0.00\\
   && No group &1.00 &1.00  &0.00 & 1.00&0.99 &0.03   \\
     \hline
0.25   &10& Ours &1.00 &1.00  &0.00 &1.00 &1.00 & 0.00  \\
   && No group &1.00 & 1.00 &0.00 &1.00 &1.00 & 0.00  \\
     \hline
0.5   &5& Ours & 1.00& 0.99 &0.03 &1.00 &0.99 & 0.03  \\
   && No group &0.80 & 0.99 &0.04 & 0.93& 0.99&0.03   \\
     \hline
0.5  &10& Ours & 1.00&  0.99&0.03 &1.00 &0.99 &0.03   \\
   & & No group & 0.80& 1.00 &0.00 &0.97 &0.99 & 0.03  \\
\hline
  \end{tabular}
}
\end{table}
       
We compare the variable selection performance of the two methods in terms of the sensitivity, specificity and false discovery rates (FDR). We also evaluate the estimation accuracy of the regression coefficients via $\mathcal{L}_1$ error defined by $\sum_{i=1}^p |\hat{\beta}_i-\beta_i|$ averaged over the repetitions. Additionally, for our proposed method, we also evaluate the consistency in detecting the true number of clusters and measure the clustering accuracy via normalised mutual information (NMI) \cite{strehl2002cluster}. 

\begin{table}[!htb]
\caption{\textbf{Estimation accuracy of the regression coefficients and clustering consistency.} For varying $\rho\in\{0.25, 0.5\}$ and censoring rate $p\in\{5\%, 10\%\}$.}\label{tab:performance2}
\small\setlength\tabcolsep{14pt}
\scalebox{0.5}{
  \begin{tabular}{cccccccccc}
    \hline
      \multicolumn{1}{c}{} &
      \multicolumn{1}{c}{} &
      \multicolumn{1}{c}{} &
      \multicolumn{3}{c}{$n_1 = n_2= 100$} &
      \multicolumn{3}{c}{$n_1 = n_2= 200$}\\
    \hline
   
     \multicolumn{1}{c}{$\rho$ } &
     \multicolumn{1}{c}{$p$ } &
     \multicolumn{1}{c}{Method } &
      \multicolumn{1}{c}{$\mathcal{L}_1$ error} &
      \multicolumn{1}{c}{$\hat{K}$} &
      \multicolumn{1}{c}{NMI} &
      \multicolumn{1}{c}{$\mathcal{L}_1$ error} &
      \multicolumn{1}{c}{$\hat{K}$} &
      \multicolumn{1}{c}{NMI} & \\
    \hline
 0.25  & 5& Ours &18.02 & 2 & 0.79&5.88 &2 &  0.73 \\
   & & No group &71.63 & $-$ &$-$ &74.34 & $-$& $-$ \\
     \hline
 0.25 & 10 &   Ours &19.64 & 2 &0.76 & 6.93&2 &0.78   \\
  & &  No group & 71.93&$-$  &$-$ &78.78 &$-$ & $-$  \\
     \hline
 0.5 & 5&   Ours & 25.27& 2 &0.79 &6.54 &2 &0.82  \\
 & &   No group & 72.00& $-$ &$-$ &74.29 &$-$ &$-$  \\
     \hline
 0.5  & 10 &  Ours &26.77 &2  &0.76 &6.61 &2 &0.71   \\
  & &  No group & 71.99& $-$ &$-$ & 74.34&$-$ &   $-$\\
     \hline
  \end{tabular}
}
\end{table}

In terms of variable selection accuracy, the gains from the proposed methodology over the Cox proportional hazards model without grouping are minimal for lower levels of correlation in the design matrix. Nonetheless, the proposed methodology tends to ensure higher sensitivity and specificity, as well as lower false discovery rates across the various simulation setups, as shown in Table \ref{tab:performance1}. The advantages become more pronounced as the correlation in the design matrix increases. The primary benefit of the proposed methodology is demonstrated through improved estimation accuracy of the regression coefficients, measured via $\mathcal{L}_1$ error, as presented in Table \ref{tab:performance2}. Additionally, Table \ref{tab:performance2} shows that the $\mathcal{L}_1$ error of the estimated regression coefficients decreases sharply with sample size for the proposed method—a phenomenon not observed with the Cox proportional hazards model with no groups. Regarding clustering accuracy, as shown in Table \ref{tab:performance1}, the number of clusters is consistently estimated via the proposed method across different simulation setups, and the normalized mutual information varied between $70-80\%$, indicating favorable clustering performance.\\
\begin{table}[!htb]
\caption{\textbf{Lung cancer data.} Cluster specific variable significance across $\hat{K} = 3$ clusters. }\label{tab:lung}
\small\setlength\tabcolsep{14pt}
\scalebox{0.5}{
  \begin{tabular}{cccccccc}
    \hline
   
     \multicolumn{1}{c}{Cluster } &
     \multicolumn{1}{c}{\texttt{age} } &
     \multicolumn{1}{c}{\texttt{ph.ecog}} &
      \multicolumn{1}{c}{\texttt{ph.karno}} &
      \multicolumn{1}{c}{\texttt{pat.karno}} &
      \multicolumn{1}{c}{\texttt{meal.cal}} &
      \multicolumn{1}{c}{\texttt{wt.loss}} \\
    \hline
1  & \cmark&\cmark &\xmark&\xmark  &\xmark & \xmark \\
\hline
2  & \xmark&\cmark &\xmark&\xmark  &\xmark & \xmark  \\
\hline
3  & \xmark&\cmark &\xmark&\xmark  &\xmark & \cmark  \\
\hline
\end{tabular}
}
\end{table}

\noindent\textbf{Case studies.} We next focus on 
two publicly real data sets \citep{survival}. \textbf{(1) Lung cancer data.} We first consider a clinical trial data on $n=167$ patients with advanced lung cancer, that includes information on survival time, censoring status, and $p=6$ covariates such as \texttt{age}, ECOG performance score (\texttt{ph.ecog}),  Karnofsky performance scores rated by both physician (\texttt{ph.karno})  and patient (\texttt{pat.karno}), average calorie intake (\texttt{meal.cal}) and weight loss during the study (\texttt{wt.loss}). The censoring rate is $28\%$. The proposed methodology identifies $\hat{K} = 3$ latent clusters and cluster-specific variable significance (Table \ref{tab:lung}), and ensures a $9\%$ increase in predictive accuracy compared to the Bayesian Cox proportional hazard model with no groups. \noindent\textbf{(2) Diabetic retinopathy data.} The diabetic retinopathy dataset comprises clinical data on $n=394$ patients with diabetes, focusing on the development of diabetic retinopathy. The dataset includes survival time, censoring status, and $p=4$ predictors such as \texttt{age}, a factor \texttt{eye} with levels left and right coded as $1$ and $0$ respectively, a factor \texttt{laser} with levels xenon and argon coded as $1$ and $0$ respectively, and \texttt{treatment} type coded as $1$ and $0$ respectively.  The censoring rate is $39\%$. The proposed methodology identifies $\hat{K} = 4$ latent clusters and cluster-specific variable significance (Table \ref{tab:diabetic}), and ensures $10\%$ increase in predictive accuracy compared to the Bayesian Cox proportional hazard model with no groups.

\begin{table}[!htb]
\caption{\textbf{Diabetic retinopathy data.} Cluster specific variable significance across $\hat{K} = 4$ clusters. }\label{tab:diabetic}
\small\setlength\tabcolsep{8pt}
\scalebox{0.5}{
  \begin{tabular}{ccccccccccccc}
    \hline
   
     \multicolumn{1}{c}{Cluster } &
     \multicolumn{1}{c}{\texttt{laser} } &
     \multicolumn{1}{c}{\texttt{age}} &
      \multicolumn{1}{c}{\texttt{eye}} &
      \multicolumn{1}{c}{\texttt{treatment}} &
      \multicolumn{1}{c}{} &
      \multicolumn{1}{c}{Cluster } &
     \multicolumn{1}{c}{\texttt{laser} } &
     \multicolumn{1}{c}{\texttt{age}} &
      \multicolumn{1}{c}{\texttt{eye}} &
      \multicolumn{1}{c}{\texttt{treatment}} &\\
    \hline
1  & \xmark&\cmark &\xmark&\xmark &&3  & \xmark&\cmark &\xmark&\cmark  \\
\hline
2  & \cmark&\cmark &\cmark&\cmark& &4  & \xmark&\cmark &\xmark&\xmark \\
\hline
\end{tabular}
}
\end{table}

\section{Conclusion}
In this article, we introduced a flexible mixture of Cox proportional hazard models that can identify latent clusters in the population of interest and determine cluster-specific variable significance. The latent clusters can potentially help practitioners identify varying risk profiles in non-homogenous populations of interest and devise targeted treatment schemes for disease management. The proposed methodology is complemented with an efficient computational algorithm. 
\clearpage
\bibliographystyle{plainnat}
\bibliography{paper-ref.bib, additionalrefs.bib}

\end{document}